\documentclass[letterpaper, 10 pt, conference]{ieeeconf}

\IEEEoverridecommandlockouts

\usepackage{float}
\usepackage{xcolor}
\usepackage{outlines}
\usepackage{graphicx}
\usepackage{amsmath,amssymb,amsfonts}
\newtheorem{assumption}{Assumption}
\newtheorem{definition}{Definition}
\usepackage{mathtools}
\usepackage[hidelinks]{hyperref} % [pagebackref=true]
\usepackage{cite}
\usepackage{mathrsfs}
\usepackage{subfigure}

\title{\LARGE \bf
Robust Model Predictive Control of Fast Lithium-ion Battery Pretreatment for Safe Recycling}

\author{Meng Yuan, Adam Burman, and Changfu Zou% <-this % stops a space
\thanks{This work is supported by the European Union (EU)-funded Marie Sklodowska-Curie Actions (MSCA) Postdoctoral Fellowship (under grant No. 101110832) and Chalmers University of Technology through the Energy Area of Advance.}% 
\thanks{Meng Yuan, Adam Burman, and Changfu Zou are with the Department of Electrical Engineering, Chalmers University of Technology, 412 96 Gothenburg, Sweden. (Emails: {\tt\small meng.yuan@chalmers.se; adambu@chalmers.se; changfu.zou@chalmers.se})
}%
}

\begin{document}

\maketitle
\thispagestyle{empty}
\pagestyle{empty}

%%%%%%%%%%%%%%%%%%%%%%%%%%%%%%%%%%%%%%%%%%%%%%%%%%%%%%%%%%%%%%%%%%%%%%%%%%%%%%%%
\begin{abstract}

The proper disposal and repurposing of end-of-life electric vehicle batteries are critical for maximizing their environmental benefits. This study introduces a robust model predictive control (MPC) framework designed to optimize the battery discharging process during pre-treatment, ensuring both efficiency and safety. The proposed method explicitly incorporates temperature constraints to prevent overheating and potential hazards. By leveraging a control-oriented equivalent circuit model integrated with thermal dynamics, the MPC algorithm dynamically adjusts the discharging profile to maintain safe operating temperatures. Additionally, the robust controller is designed to account for model mismatches between the nonlinear battery dynamics and the linearized model, ensuring reliable performance under varying conditions. The effectiveness of this approach is demonstrated through simulations comparing the robust MPC method with conventional discharging strategies, including constant current-constant voltage (CC-CV) and constant current-constant temperature (CC-CT) methods. Results indicate that the robust MPC framework significantly reduces discharging time while adhering to safety constraints, offering a promising solution for the recycling and second-life applications of lithium-ion batteries.

\end{abstract}

%%%%%%%%%%%%%%%%%%%%%%%%%%%%%%%%%%%%%%%%%%%%%%%%%%%%%%%%%%%%%%%%%%%%%%%%%%%%%%%%
\section{Introduction}

% Importance of battery recycling
Electric vehicles (EVs) have been widely adopted and are gaining popularity due to their potential to reduce carbon emissions. However, improper disposal of end-of-life batteries, such as placing them in landfills, diminishes these environmental benefits and causes new problems, such as soil contamination and water pollution \cite{tao2021second, shahjalal2022review}. In EVs, batteries are typically retired when their capacity falls below a certain threshold, usually 70\% or 80\% of their original capacity \cite{wood2011investigation}. Instead of discarding these batteries, they can be repurposed as energy storage systems, particularly when paired with renewable energy sources, for second-life use.

% Difficulties in pre-treatment of batteries. 
The rising demand for batteries underscores an increasing focus not only on the production of lithium-ion batteries but also on extending their lifespan through advanced battery management systems and recycling after second-life applications \cite{wang2011optimal, akar2016energy, yang2022modelling}. Lithium-ion batteries contain valuable materials, including lithium and high-grade copper and aluminum, and may also contain cobalt and nickel, depending on the active material. To prevent a future shortage of cobalt, nickel, and lithium and to ensure the sustainable life-cycle of batteries, it is essential to develop efficient recycling technology for lithium batteries \cite{harper2019recycling}. However, to ensure the safety of battery recycling, particularly during the disassembly of battery systems and mechanical processes, including crushing, sorting, and sieving processes, the first and fundamental step is to discharge each battery cell thoroughly \cite{kim2021comprehensive, shi2024current}. In principle, the lower residual charge capacity in the cell implies a safer recycling process. 

% Importance of controlling temperature
At the same time, controlling temperature during battery discharge is critical since it relates to safety and stable discharge performance. On the one hand, maintaining a relatively high temperature during discharge can be beneficial, as experimental results indicate that higher battery temperatures during discharge can result in lower residual energy after a complete cycle \cite{mondal2024pretreatment}. On the other hand, a too-high temperature during discharging can lead to safety issues. If a lithium-ion battery is not effectively monitored during discharge, an external short circuit may occur, leading to a rapid rise in temperature. When the battery temperature reaches 77 to 121$^{\circ}$C, venting and electrolyte leakage may occur \cite{chen2021review}. Once thermal runaway begins, the internal structure of the battery changes, making it unsafe to operate or discharge \cite{kong2021strategies}. At this stage, the battery must be handled according to predefined safety measures, as a high current through a damaged battery can result in an explosion. 

Thus, selecting a temperature that is high yet remains within safe limits is crucial. For battery charging, the CC-CT (constant current-constant temperature) method is already a mature approach to maintaining temperature during charging \cite{sabarimuthu2023multi}. This naturally leads us to consider using the CC-CT method for battery discharging as well, as to control temperature simultaneously.

Then the primary research problem in battery discharging becomes balancing the competing objectives of achieving a lower residual charge capacity and a faster discharge rate. This challenge is made more difficult by safety requirements, which require that temperatures stay within specific limits throughout discharge, and by the uncertainty of initial battery capacities. These constraints often result in slower discharge rates and frequent switching between constant current (CC) and constant voltage (CV) stages, further extending the discharge duration. Optimizing this process to meet both speed and safety objectives is therefore a critical area of focus in both academic and industrial research.

Given the need to balance discharge speed and temperature safety, model predictive control (MPC) is an effective approach for optimizing battery discharging. MPC is a constraint-based control method that optimizes control actions over a finite prediction horizon. It computes the optimal control strategy by minimizing a cost function that penalizes deviations in both future states and control inputs, ensuring that states and control signals remain within predefined constraints \cite{yuan2019error, yuan2023safety}. MPC is well-suited for battery discharging, as it can penalize both process time and temperature, enabling fast discharge while avoiding temperature violations to ensure safety. In \cite{de2020lithium}, MPC is applied with a low-level equivalent circuit model (ECM) to manage cell-level charging of lithium-ion batteries.%  (LIBs)

However, to the best knowledge of the authors, there is limited research on using advanced controllers to battery discharge for safe recycling purpose, with the constant current method still prevalent in pre-treatment processes \cite{traub2016calculation}. There is significant potential in exploring methods such as model predictive control to optimize discharge time while ensuring safety during the discharging process.

Therefore, the main contribution of this work is presenting a discharging framework of lithium batteries based on a robust MPC algorithm to accelerate the entire pre-treatment process time while ensuring a given temperature constraint is not violated for safety purposes. The resulting discharging profile is compared with three benchmark control strategies namely, CC-CV, CC-CT, and dynamic programming (DP). 

The remainder of this article is organized as follows: In Section II, the cell-level equivalent circuit model is integrated with thermal modeling to serve as the control-oriented model, and the modeling procedure is described. Section III presents the formulation of conventional discharging controllers, including CC-CV, CC-CT, DP, and the proposed robust MPC control. The discharging results of a cell battery using the proposed method, along with benchmark controllers, are presented in Section IV. Finally, Section V concludes the article.

Notation: The real and natural numbers are denoted as $\mathbb{R}$ and $\mathbb{N}$, respectively. Given two integers $a, b \in \mathbb{N}$, the integer range is denoted by $\mathbb{N}_{[a,b]} \triangleq \{i \in \mathbb{N} \mid  a \leq i \leq b\}$. The symbol $x(i|k)$ stands for the predicted value of $x$ based on the measurement at time $k$. The measurement value of $x$ at time instant $k$ is represented as $x(k)$. A diagonal matrix with main diagonal elements $(a_{1},\cdots, a_{n})$ is denoted by $\text{diag}(a_{1},\cdots, a_{n})$.

\section{Modeling}

In this section, the model for controller design purposes is presented. An equivalent circuit model with a zero-dimension lumped mass heat equation is adopted as the control-oriented model. 

\subsection{Electrothermal model of battery} % 

\begin{figure}
    \centering
    \includegraphics[width=0.9 \linewidth]{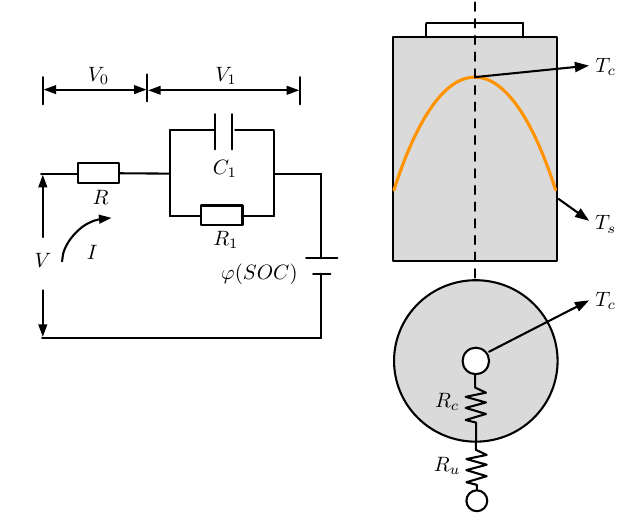}
    \caption{Illustration of electrothermal modeling of a lithium-ion battery \cite{zou2017electrothermal}.}
    \label{fig:ECM_diagram}
\end{figure}

The schematic diagram of the system is shown in Fig.~\ref{fig:ECM_diagram}. The electrical behavior of the battery cell is modeled using the ECM, which consists of a voltage source representing the open-circuit voltage, a resistor $R$ for internal resistance, and a parallel resistor-capacitor (RC) pair to capture the dynamic response of the system.

When modeling a battery using an ECM, multiple parallel RC pairs in series can be used to more accurately capture the time dynamics \cite{li2021control}. This approach is especially beneficial for improving model accuracy when the system experiences rapid changes in input current. If the input current varies slowly or remains relatively stable, the system enters a quasi-steady state, making transient effects negligible compared to the overall operation.

Balancing model accuracy with computational efficiency is crucial when considering model complexity, parameter estimation, and computational power. In this work, the input current changes at a relatively low rate during the discharge cycle, indicating that a single RC pair is sufficient to capture the system dynamics.

The states used to describe the electrical model are state of charge $\text{SoC}(t)$, voltage over the RC pair $V_1(t)$, surface temperature $T_s(t)$, and core temperature $T_c(t)$. Then, the electrical and thermal governing equations for a lithium-ion battery are given as \cite{zou2017electrothermal}:
\begin{subequations}\label{eq:state_equations}
    \begin{align}
\frac{d\text{SoC}(t)}{dt} &= -\frac{I(t)}{3600 C_n}, \label{eq:state_equations_a}\\
\frac{dV_{1}(t)}{dt} &= -\frac{V_{1}(t)}{R_{1} C_{1}} + \frac{I(t)}{C_{1}}, \label{eq:state_equations_b}\\
\frac{dT_s(t)}{dt} &= \frac{T_f - T_s(t)}{R_u C_s} - \frac{T_s(t) - T_c(t)}{R_c C_s}, \label{eq:state_equations_c} \\
\frac{dT_c(t)}{dt} &= \frac{T_s(t) - T_c(t)}{R_c C_c} + \frac{I(t)}{C_c}\big{[}V_{1}(t) + R_0 I(t)\big{]}, \label{eq:state_equations_d}
\end{align}
\end{subequations}
 where $I(t)$ is the input current, $C_n$ is the nominal capacity of the battery cell, $R_u$ and $R_c$ are the heat convection and conduction resistances, respectively, $T_f$ is the ambient temperature, and $C_s$ and $C_c$ are the surface and core heat capacity, respectively.

The measured system outputs include surface temperature $T_{s}(t)$ and the terminal voltage $V(t)$. The expression of $V$ is a nonlinear combination of system states, and its mathematical representation is given as
\begin{equation}
    V(t) \triangleq \varphi (\text{SoC}(t)) + V_{1}(t) + R_{0}I(t),
_{}\end{equation}
where $\varphi(\cdot)$ represents the open circuit voltage and is a nonlinear function of $\text{SoC}$.

\subsection{State of energy computation}

In this work, one of the key metrics used to evaluate various discharge strategies is the time required for complete discharge. The proposed method for determining full battery discharge involves measuring the extracted energy and comparing it to the nominal energy capacity. The extracted energy is calculated as follows:

\begin{equation}
    E_{o}(t) = \int_{0}^{t} V(\tau)I(\tau) d\tau.
\end{equation}

The nominal energy $E_{n}$, representing the total energy stored in the battery, is determined by fully discharging and charging it to full capacity. The conventional CC-CV method with a low current can be used for this process \cite{wei2020unbiased}: 

\begin{equation}
    E_{n} = \int_{0}^{T_{f}} V(t)I(t) dt,
\end{equation}
where $T_{f}$ is the time taken to charge the depleted battery to full capacity. Then, the state of energy (SoE) is defined as
\begin{equation}
    \text{SoE}(t) = 1 - E_{o}(t)/E_{n},  
\end{equation}
which can be discretized as:
\begin{equation}\label{eq:SoE_discrete}
    \text{SoE}(k) = \text{SoE}(k-1) - \frac{\eta V(k-1)I(k-1)\Delta t}{E_{n}},
\end{equation}
where $\eta$ denotes the energy efficiency of the battery, and $\Delta t$ represents the sampling time of the system \cite{wei2020unbiased}. Considering that the battery used for recycling was previously installed in an EV with a mature battery management system, the following assumption is made.

\begin{assumption}
    Based on the information from a mature energy management system, the value of $E_{n}$ is assumed to be known when implementing the discharging controller for battery recycling.
\end{assumption}

Let the system states be \( x(t) \triangleq [\text{SoC}(t), \allowbreak V_{1}(t), \allowbreak T_{s}(t), \allowbreak T_{c}(t)]^{\top} \), the input be \( u(t) \triangleq I(t) \), \( y(t) \triangleq [\text{SoE}(t), \allowbreak T_{c}(t)]^{\top} \) as the system outputs to be controlled, and \( z(t) \triangleq [\allowbreak T_{s}(t), \allowbreak V(t)]^{\top} \) as the measurable system outputs. The compact form of the battery model is represented as:
\begin{subequations}\label{eq:sys_compact_con}
\begin{align}
    \dot{x}(t) & = f_{c}(x(t),u(t)), \\
    y(t) & = h_{c}(x(t),u(t)),\\
    z(t) & = g_{c}(x(t),u(t)).
\end{align}
\end{subequations}

To implement a digital controller for real-time control, the continuous-time model \eqref{eq:sys_compact_con} can be discretized as follows:
\begin{subequations}\label{eq:sys_compact_dis}
    \begin{align}\label{eq:sys_compact_dis_state}
        x(k+1) =\:& f(x(k),u(k)), \\
    % \end{equation}
    % \begin{equation}
        y(k) =\:& h(x(k),u(k)), \\
    % \end{equation}
    % \begin{equation}
        z(k) =\:& g(x(k),u(k)).
    \end{align}
\end{subequations}
where $u(k)\approx u(k\cdot \Delta t)$, $k \in \mathbb{N}$.

For controller design purposes, the battery model in \eqref{eq:sys_compact_dis_state} is linearized at different operating points as:
\begin{subequations}\label{eq:sys_linear}
    \begin{align}
        x(k+1) & = Ax(k) + Bu(k) + w(k), \label{eq:state_linear}\\
        y(k) & = h(x(k),u(k)),
    \end{align}
\end{subequations}
where $w\in \mathcal{W} \subseteq \mathbb{R}^{5}$ is an unknown but bounded lumped disturbance that includes model mismatch due to linearization and parameter variation. Here, the nonlinear output is a function of the system states and input. Linearizing this output could lead to a loss of accuracy, particularly in capturing the inherent nonlinearities of batteries that are essential for accurate SoE estimation. 

To control the core temperature, it is necessary to design an estimator to acquire the system state. In this work, a Kalman filter serves as the state estimator, based on the measured output $z(k)$. The process noise covariance and sensor noise covariance are the two design parameters for configuring this estimator.

\section{Controller formulation}

In this section, we briefly discuss the benchmark controllers, namely CC-CV, CC-CT, and DP. Then, we provide a detailed formulation of the proposed controller.

\subsection{Conventional approach}

\subsubsection{CC-CV method}

Constant current-constant voltage (CC-CV) discharging remains the most widely used method for battery pre-treatment. In this method, the battery is initially discharged at a constant current until it reaches a predefined cut-off voltage. This is followed by a constant voltage phase, during which conventional feedback controllers, such as PI controllers, maintain the voltage until the cut-off current is reached \cite{mondal2024pretreatment}. The battery is considered fully discharged once the cut-off current value is achieved \cite{wei2020unbiased}.

\subsubsection{CC-CT method}

The CC-CT controller operates similarly to the CC-CV controller, except that the constant voltage phase is replaced with a constant temperature phase. Since high temperatures during discharge can reduce residual energy, selecting the temperature setpoint involves a trade-off between minimizing remaining energy and ensuring safety.

\subsubsection{Dynamic programming}

When the battery system model is accurate, an optimization-based method can be used to compute the control input \cite{xu2019fast}. In this work, dynamic programming is employed to calculate the optimal discharging current profile, aiming to minimize the SoE as quickly as possible while adhering to the temperature constraint. This approach provides an offline solution to the optimal discharging problem, with the optimal control input determined by solving the following optimization problem:

    \begin{subequations}\label{problem:RAMPC}
    \begin{align}
    & \underset{u}{\min} \;
    \begin{cases}
    \begin{multlined}
         N_{h} + w_{1} |\text{SoE}(N_{h}|k)| \\
          + w_{2} \sum\limits_{i=0}^{N_{h}} u(i|k),
    \end{multlined}
			 & \text{if } \text{SoE}(i|k) \geq 0 \\
    w_{3} N_{f} + w_{4} \sum\limits_{i=0}^{N_{f}} u(i|k), & \text{if } \text{SoE}(N_{f}|k) <0 %\exists t_f: x_5(t_f) \leq 0   
	\end{cases} \\
			%%%%%%%%%%%%%%%%
	 \intertext{subject to}  & x(i+1|k) = f(x(i|k),u(i|k)), \\
                & y(i|k) = h(x(i|k),u(i|k)), \\
			&  T_{c}(i|k) \leq T_{\max}, \\
            & 0 \leq u(i|k) \leq u_{\max}, \\ % i = 0,\cdots, N_{h}, 
			& x(0|k) = [\text{SoC}(k), V_{1}(k), T_{s}(k), T_{c}(k)]^{\top},
		\end{align}
	\end{subequations} % x(k) = [1,0,20,20,1]
where $T_{\max}$ is the upper bound of the core temperature; $u_{\max}$ is the upper bound of the control input.

\subsection{Proposed approach}

\begin{figure}
    \centering
    \includegraphics[width=\linewidth]{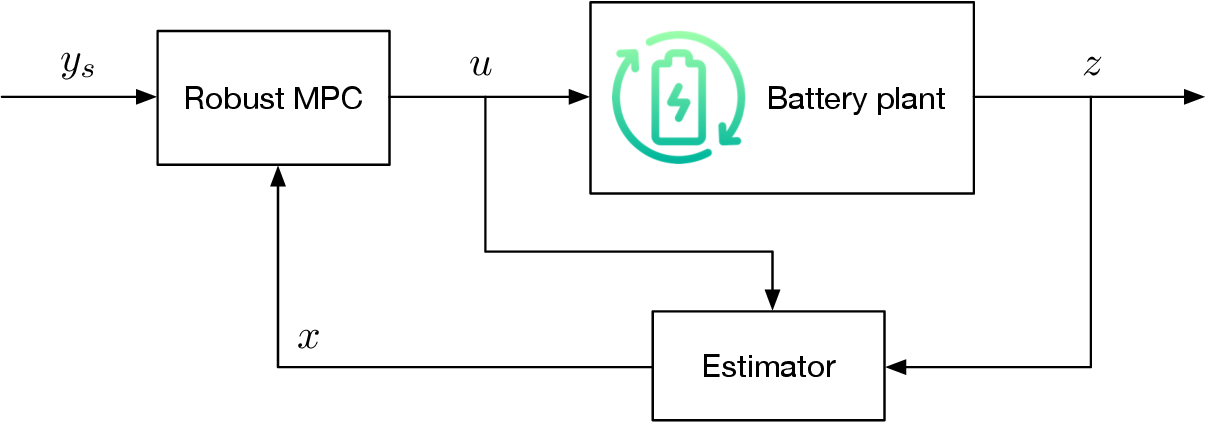}
    \caption{Proposed control structure for battery discharging.}
    \label{fig:whole_struc}
\end{figure}

In this work, our goal is to achieve rapid battery discharge while consistently meeting temperature constraints. To this end, a robust MPC is employed as the controller, and the entire control structure is illustrated in Fig.~\ref{fig:whole_struc}.

We begin by defining a polyhedral set that encapsulates the state and input constraints required during the discharging process as follows:
\begin{equation}
    (x(k),u(k))\in \mathcal{L}, \label{eq:cons_ori}
\end{equation}
where $ \mathcal{L} \triangleq \{q \in \mathbb{R}^{6}: A_{q} q \leq b_{q}\}$. For the linearized system \eqref{eq:state_linear}, its nominal model is given as:
\begin{equation}
    \bar{x} = A \bar{x} + B\bar{u},
\end{equation}
where $\bar{x}$ and $\bar{u}$ are the nominal state and input, respectively.

Let the error between the linearized and nominal system be $e\triangleq x - \bar{x}$, and design the control action as:
\begin{equation}\label{eq:control_u}
    u(k) = \bar{u}(k) + K(x(k) - \bar{x}(k)),
\end{equation}
where $K$ is the feedback control gain that is selected to ensure $(A+BK)$ is a Hurwitz matrix. Then, the dynamics of the error system between the nominal and linearized systems become:
\begin{equation}\label{eq:sys_err}
    e(k+1) = A_{K}e(k) + w(k),
\end{equation}
where $A_{K} = (A + BK)$. In order to design the robust MPC as the proposed discharging controller, the notation of a robust positively invariant (RPI) set is introduced first. 
\begin{definition}
    For the uncertain system \eqref{eq:sys_err}, the set $\mathcal{R}$ is a robust positively invariant if $A_{K}\mathcal{R}\oplus \mathcal{W} \subseteq \mathcal{R}$.
\end{definition}

Then, if $x(0) \in \bar{x}(0) \oplus \mathcal{R}$, a tighter constraint for the state and input can be formulated as follows:
\begin{equation}
    (\bar{x}(k), \bar{u}(k)) \in \bar{\mathcal{L}}, \label{eq:cons_tighter}
\end{equation}
with $\bar{\mathcal{L}} \triangleq \mathcal{L} \ominus(\mathcal{R}\times K\mathcal{R})$. By ensuring the tighter constraint satisfaction of \eqref{eq:cons_tighter} using control input \eqref{eq:control_u}, the original constraint \eqref{eq:cons_ori} can be ensured and the uncertain system controlled is robustly admissible \cite{limon2010robust}. The computation of the minimal robust invariant set $\mathcal{R}$ can be conducted based on the methods in \cite{rakovic2005invariant}. 

In the following, we will design the robust MPC to compute the control input $\bar{u}$. With a given desired output $y_{s} \triangleq [\text{SoE}^{*}, T_{c}^{*}]^{\top}$, the optimal control policy $\bar{u}$ is calculated by minimizing the following problem at each time instant $k$:

\begin{subequations}\label{eq:MPC_problem}
	\begin{align}
		& \underset{\bar{u}, \bar{x}(0|k)}{\min} \sum_{i=0}^{N-1}\left(\left\Vert y_{s} - y(i|k) \right\Vert _{Q}^{2}  +\left\Vert \bar{u}(i|k)\right\Vert _{R}^{2}\right), \\
		\intertext{subject to}\; & \bar{x}(i+1|k)=A \bar{x}(i|k)+B \bar{u}(i|k),\label{eq:MPC_system} \\
            & y(i|k) = h(x(i|k),u(i|k)), \\
        & (\bar{x}(i|k), \bar{u}(i|k)) \in \bar{\mathcal{L}}, \\
        & -1<= u(i+1|k) - u(i|k) <=1,  \, i\in \mathbb{N}_{[0,N-1]}, \\
         & \bar{x}(0|k) \in x(0) \oplus (-\mathcal{R}). \label{eq:x0_MPC} 
	\end{align} %
\end{subequations}

At each time instant $k$, the control input is given as $u(k) = \bar{u}^{*}(0|k) + K(x(k) - \bar{x}^{*}(0|k))$. The cost function in \eqref{eq:MPC_problem} penalizes the SoE while taking account of the input change at each time step.

\section{Results and Discussion}

\begin{figure}[htbp]
		\subfigure[]{
			\includegraphics[width = 0.85\columnwidth]{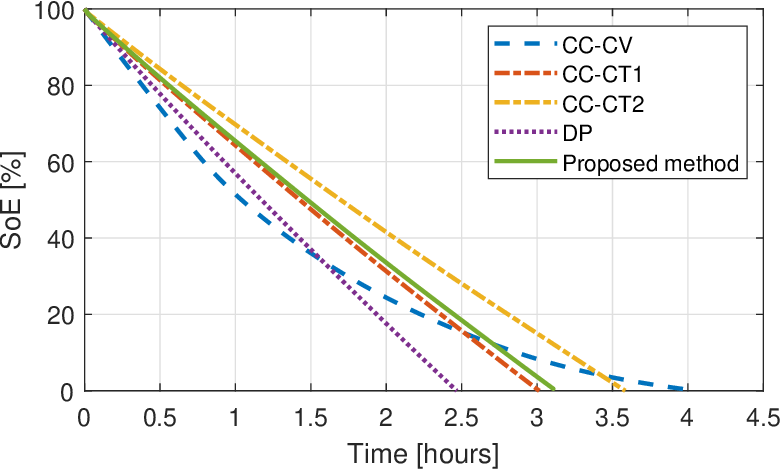}\label{fig:result_soe}
		}
		\subfigure[]{
			\includegraphics[width = 0.85\columnwidth]{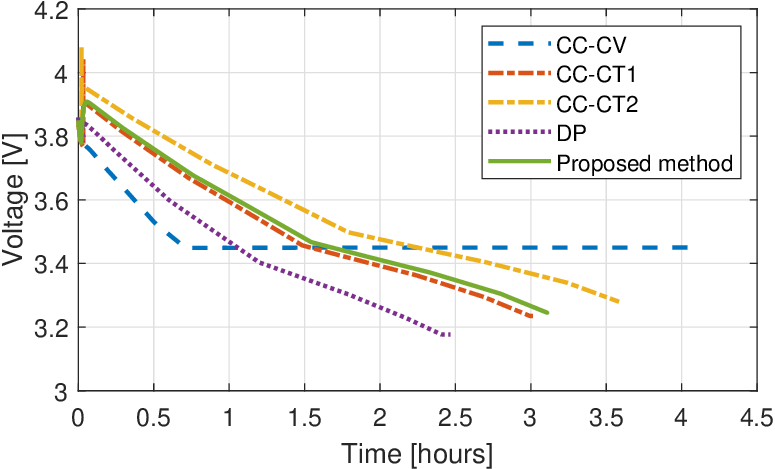}\label{fig:result_vol}
		}
        \subfigure[]{
			\includegraphics[width = 0.85\columnwidth]{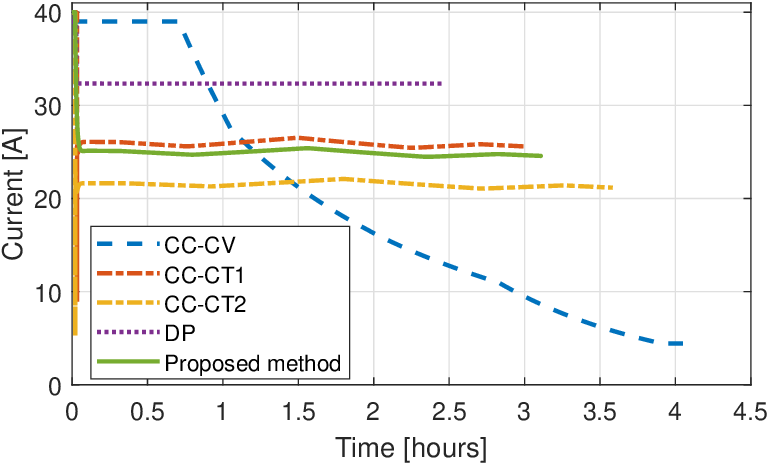}\label{fig:result_cur}
		}
        \subfigure[]{
			\includegraphics[width = 0.85\columnwidth]{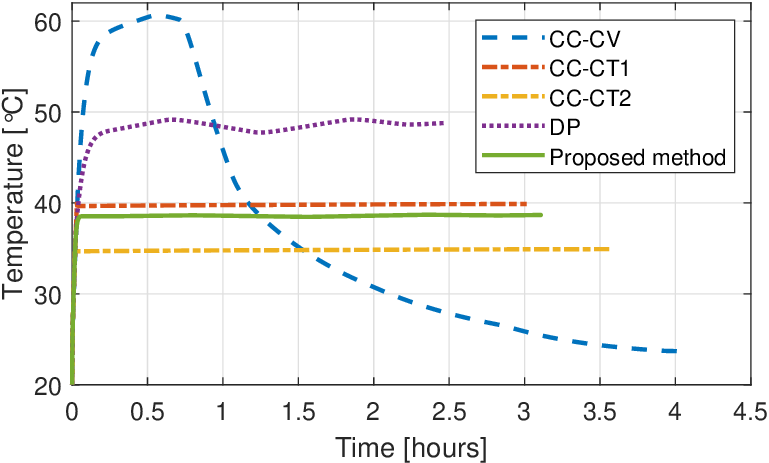}\label{fig:result_temp}
		}
		\caption{Battery discharging performance based on different methods: (a) SoE; (b) terminal voltage; (c) input current; (d) core temperature.}
		\label{fig:results_four}
\end{figure}  

To demonstrate the effectiveness of the proposed controller, battery discharging is simulated using the CC-CV, CC-CT, dynamic programming, and the proposed robust MPC methods with a high-fidelity Simscape battery model in MATLAB. The performance of each controller is evaluated based on the time taken to complete the discharge and the satisfaction of temperature constraints. 

Given that a higher battery temperature during discharge can enhance energy extraction and reduce residual energy, the core temperature tolerance is set to $40^{\circ}$C in this study \cite{mondal2024pretreatment, chen2014accurate}. For the CC-CT controller, an additional test was conducted with a temperature tolerance of $35^{\circ}$C, with the results referred to as CC-CT1 and CC-CT2, respectively. This comparison is intended to illustrate the trade-off between discharging time and temperature constraint satisfaction.

For a fair comparison between the different controllers, the batteries are assumed to be fully charged at the start of the simulation, with an ambient temperature of $20 ^{\circ}$C. The nominal state of energy is determined through the CC-CV discharging scheme from a fully charged battery. 

For the CC-CV controller, the initial discharging current is set as $40$ A. This constant current discharging process is followed by a constant voltage stage when the voltage drops below $3.45$ V. The entire discharging process stops when the state of energy reaches $0$. The controller gains are chosen as $K_{P} = 50$ and $K_{I} = 10$. 

For the CC-CT controller, it starts with the same $40$ A constant current phrase, followed by a temperature-maintaining process using a PI controller with parameters $K_{P} = 60$ and $K_{I} = 0.0061$. The tuning parameters for dynamic programming are chosen as $w_{1} = 10^{5}$, $w_{2} = 1/10^{5}$, $w_{3} = 10$, $w_{4} = 1/10^{5}$, $T_{max}  =40 $ and $u_{max} = 40$, with a sampling time of $20$ s. 

For the proposed controller, the robust MPC is formulated using YALMIP \cite{lofberg2004yalmip}, and the optimization is solved using the Mosek solver. The controller gains are chosen as $Q = \text{diag}(10^{4},10^{4})$, and $R = 1$.

The discharging performance using four different methods is demonstrated in Fig.~\ref{fig:results_four}, while the discharging time and constraint satisfaction results are summarized in Table.~\ref{tab:result_table}. The figures illustrate the discharging performance in terms of discharging speed, voltage and current profile, and thermal behavior.

It can be seen that the DP-based method achieves the fastest discharging time, however, the core temperature violates the constraint with the maximum temperature reaching around $50^{\circ}$C. This is mainly due to the model mismatch between the plant and the linearized model used when computing the control input. 

The CC-CV method takes the longest time for discharging and does not ensure the temperature constraint either, since temperature information is not explicitly considered in this method. For the two CC-CT-based methods, we can see that the temperature constraint is violated when the temperature reference is set as $40^{\circ}$C and the constraint can be satisfied when a lower tolerance is set. However, this comes at the cost of a longer discharging time. 

The proposed robust MPC-based method, on the other hand, prioritizes thermal safety, achieving a controlled and steady energy reduction that outperforms the slower CC-CT2. From Fig.~\ref{fig:result_cur} and Fig.~\ref{fig:result_temp}, we can see that the proposed method dynamically adjusts the current profile when the core temperature closes to the tolerance value. There is a safety margin from the temperature threshold of 40$^{\circ}$C, demonstrating effective thermal management by the proposed method.

In Table.~\ref{tab:result_table}, it can be seen that only the CC-CT2 method and the proposed method meet the temperature constraints. The proposed method, however, achieves a faster discharge time than the CC-CT2 method. Although the discharge time of the proposed method is slightly longer than that of the CC-CT1 method, it maintains a lower maximum temperature within the required limits. In summary, while the proposed method does not have the fastest discharge rate, it effectively balances discharge speed and thermal safety, making it suitable for applications that require both efficiency and safety in battery recycling.

In addition, the designed control framework can be applied to more complex battery models, such as multiphysics pseudo-two-dimensional models \cite{li2023nonlinear} or model-integrated neural network-based models \cite{huang2024minn}. For these detailed models, constraints can be imposed directly on safety-related states, such as maximum temperature and overpotential, to further increase discharge speed while maintaining safety. Although the computation time of the proposed discharging framework increases with model complexity, the improved accuracy can reduce model mismatches, resulting in less conservative outcomes. This, in turn, allows for higher temperatures during discharge, thereby reducing the overall process time.

\begin{table}
\caption{Discharging performance for different methods}\label{tab:result_table}

\centering{}%
\begin{tabular}{cccc}
\hline 
Method & Discharge time & Max. temp. & Cons. satis. \tabularnewline
\hline 
CC-CV & 4 hrs 3 mins & $60.63^{\circ}$C & No\tabularnewline
CC-CT1 & 3 hrs 1 min & $40.10^{\circ}$C & No\tabularnewline
CC-CT2 & 3 hrs 35 mins & $35.09^{\circ}$C & Yes\tabularnewline
DP & 2 hrs 28 mins & $49.18^{\circ}$C & No\tabularnewline
Proposed method & 3 hrs 6 mins & $38.69^{\circ}$C & Yes\tabularnewline
\hline 
\end{tabular}
\end{table}

\section{Conclusions}

This study presents a robust model predictive control framework for the efficient and safe discharging of lithium-ion batteries during pre-treatment. By integrating temperature constraints directly into the control strategy, the proposed method ensures that the battery temperature remains within safe limits, thereby preventing thermal runaway and potential hazards. The robust MPC approach outperforms traditional methods such as CC-CV, CC-CT, and dynamic programming by achieving faster discharge times while maintaining safety.

The simulation results demonstrate that the proposed method not only meets temperature constraints but also significantly reduces discharging time compared to conventional methods. Among the methods satisfying these constraints, the proposed method reduces discharging time by 13.28\% compared to the conventional CC-CT approach. This improvement is particularly important for recycling and second-life applications of EV batteries, as it enhances both the efficiency and safety of the pre-treatment process. % gpt checked. 

In this work, we investigate cell-level discharging. Future research could explore thermal dynamic modeling and battery discharge at the pack level. Additionally, no current loop is implemented, meaning the discharge current is generated directly from the controller’s reference. PWM-based current control could also be considered in future work.

\bibliographystyle{IEEEtran.bst}
\bibliography{ECC_ref.bib}
 
% uncomment this at the end
% \addtolength{\textheight}{-12cm}  


%%%%%%%%%%%%%%%%%%%%%%%%%%%%%%%%%%%%%%%%%%%%%%%%%%%%%%%%%%%%%%%%%%%%%%%%%%%%%%%%

%%%%%%%%%%%%%%%%%%%%%%%%%%%%%%%%%%%%%%%%%%%%%%%%%%%%%%%%%%%%%%%%%%%%%%%%%%%%%%%%

%%%%%%%%%%%%%%%%%%%%%%%%%%%%%%%%%%%%%%%%%%%%%%%%%%%%%%%%%%%%%%%%%%%%%%%%%%%%%%%%



% \section*{ACKNOWLEDGMENT}
% %%%%%%%%%%%%%%%%%%%%%%%%%%%%%%%%%%%%%%%%%%%%%%%%%%%%%%%%%%%%%%%%%%%%%%%%%%%%%%%%

% References are important to the reader; therefore, each citation must be complete and correct. If at all possible, references should be commonly available publications.

% \end{thebibliography}

\end{document}